\documentclass[reprint,NumberedRefs]{JASA}

\begin{document}

\title[Frequency-Domain Feature Modeling for HRTF Upsampling]{Exploring Frequency-Domain Feature Modeling for Head-Related Transfer Functions Magnitude Upsampling}
\author{Xingyu Chen}
\author{Hanwen Bi}
\author{Fei Ma}
\author{Sipei Zhao}
\email{sipei.zhao@uts.edu.au}

\author{Eva Cheng}
\affiliation{Centre for Audio, Acoustics and Vibration, Faculty of Engineering and Information Technology, University of Technology Sydney, Ultimo, New South Wales 2007, Australia}
 
\author{Ian S. Burnett}			
\affiliation{Faculty of Information Technology, Monash University, Clayton, Victoria 3800, Australia}


\date{November 2026} 

\begin{abstract}
Accurate upsampling of head-related transfer functions (HRTFs) from sparse measurements is crucial for personalized spatial audio rendering. 
Traditional interpolation methods, such as kernel-based weighting or basis function expansions, rely on measurements from a single subject and are limited by the spatial sampling theorem, resulting in significant performance degradation under sparse sampling.
Recent learning-based methods alleviate this limitation by leveraging cross-subject information, yet most existing neural architectures primarily focus on modeling spatial relationships across directions, whereas spectral dependencies along the frequency dimension are often modeled implicitly or treated independently.
However, HRTF magnitude responses exhibit strong local continuity and long-range structure in the frequency domain, which are not fully exploited.
This work investigates frequency-domain feature modeling by examining how different architectural choices---ranging from per-frequency multilayer perceptrons to convolutional, dilated convolutional, and attention-based models---affect performance under varying sparsity levels, showing that explicit spectral modeling consistently improves reconstruction accuracy, particularly under severe sparsity.
Motivated by this observation, a frequency-domain Conformer-based architecture is adopted to jointly capture local spectral continuity and long-range frequency correlations.
Experimental results on the SONICOM and HUTUBS datasets demonstrate that the proposed method achieves state-of-the-art performance in terms of interaural level difference and log-spectral distortion.
\end{abstract}

\maketitle

\noindent\begin{minipage}{\columnwidth}
\small
\noindent\textbf{Author-prepared manuscript (preprint).} This document is an
author-prepared version of the article, not the publisher-formatted Version of
Record. For the final published article, please consult the DOI below.\par\smallskip
\noindent Copyright 2026 Acoustical Society of America. This article may be
downloaded for personal use only. Any other use requires prior permission of
the author and the Acoustical Society of America.\par\smallskip
\noindent The following article appeared in \emph{The Journal of the Acoustical
Society of America} \textbf{160}(5), 3663--3674 (2026), and may be found at
\url{https://doi.org/10.1121/10.0047189}.
\end{minipage}\par\bigskip

\section{\label{sec:1} Introduction}
Head-related transfer functions (HRTFs) characterize the direction-dependent acoustic filtering of 
sound waves caused by the head, torso, and pinnae, and therefore serve as a fundamental component 
of spatial audio rendering~\cite{wenzel1993perceptual,keyrouz2007binaural,geronazzo2018we}. 
Because these structural features vary significantly across individuals, 
HRTFs exhibit strong person-specific characteristics~\cite{pelzer2020head}. 
Personalized spatial audio rendering requires, for each individual, 
HRTFs at a dense set of spatial directions~\cite{oberem2020experiments}.
However, acquiring such dense measurements is costly, time-consuming, and often impractical in real-world applications.
This motivates HRTF upsampling, namely, estimating dense-direction HRTFs from a limited number of sparsely measured directions.

HRTFs encode multiple perceptual features. 
Interaural time difference (ITD) is a binaural cue for sound localization~\cite{andreopoulou2017identification}, 
whereas interaural level difference (ILD) contributes to horizontal localization~\cite{mckenzie2019interaural}. 
In addition, frequency-dependent monaural spectral features play an important role in elevation and front--back localization~\cite{carlini2024auditory}.
Accordingly, many studies decompose HRTF upsampling into two sub-tasks: estimating ITD and upsampling the log-magnitude spectrum, followed by minimum-phase reconstruction to obtain the time-domain impulse response, referred to as the head-related impulse response (HRIR)~\cite{mcmullen2022machine,fantini2025survey}.
As ITD estimation already achieves high accuracy~\cite{zhao2025head}, this work focuses on log-magnitude HRTF upsampling.

Traditional HRTF upsampling methods primarily focus on modeling spatial relationships among measured directions, following early spatial weighting
and panning concepts developed for virtual sound source positioning
\cite{pulkki1997virtual}.
These methods estimate unmeasured directions using only the target subject's
available measurements and can be broadly categorized into distance-weighted
interpolation methods and basis-function decomposition methods.
The former, including bilinear interpolation~\cite{begault20003}, spherical-triangle interpolation~\cite{freeland2004interpositional}, and barycentric interpolation~\cite{hartung1999comparison,cuevas20193d}, estimates target HRTFs by computing weighted combinations of neighboring measurements.
The latter, such as spherical harmonic decomposition~\cite{zotkin2009regularized, ahrens2012hrtf,porschmann2019directional,arend2021assessing} and spatial principal component analysis~\cite{xie2012recovery, zhang2020modeling}, represents HRTFs as linear combinations of spatial basis functions, with interpolation achieved by estimating the corresponding spatial coefficients from the measured directions.
Although effective under dense measurement conditions, these traditional approaches do not exploit inter-subject relationships and suffer substantial accuracy degradation under sparse sampling, as fundamentally constrained by the spatial sampling theorem.

Recent learning-based methods have improved HRTF upsampling by exploiting inter-subject relationships enabled by multi-subject datasets~\cite{watanabe2014dataset, sridhar2017database, armstrong2018perceptual, brinkmann2019hutubs, engel2023sonicom}.
Existing approaches can be broadly categorized into coordinate-based models and
sparse-to-dense regression models. Coordinate-based models represent HRTFs as
continuous functions of spatial coordinates conditioned on subject-specific
latent representations, whereas sparse-to-dense regression models directly
predict dense-direction HRTFs from a limited set of sparse measurements.
The former includes early autoencoder-based models conditioned on source
direction~\cite{ito2022head,ito2025spatial}, followed by neural-field methods~\cite{zhang2023hrtf,lee2023global,masuyama2024niirf} inspired by neural radiance field (NeRF)~\cite{mildenhall2021nerf}.
Building on this framework, subsequent extensions incorporate additional spatial or
physical priors. 
These include graph neural networks for explicit spatial
correlation modeling~\cite{hu2025graph}, retrieval-augmented mechanisms that
leverage similar subjects' HRTFs from a database, such as Retrieval-Augmented
Neural Field (RANF)~\cite{masuyama2025retrieval},
and physics-informed neural networks that incorporate the Helmholtz equation into the training loss~\cite{ma2023spatial}.
The latter directly maps a limited set of sparse measurements to dense-direction
HRTFs without explicit coordinate conditioning.
Different neural network architectures have been investigated within this formulation.
Representative examples include per-frequency multilayer perceptron (MLP) models~\cite{zhao2025head}, variational autoencoder-based
models~\cite{zurale2023spatial}, and generative adversarial networks
(GANs)~\cite{hogg2024hrtf}.
Within this category, several methods further formulate HRTFs as image-like representations over direction and frequency.
For example, Jiang et al.~\cite{jiang2023modeling} arranged HRTFs on
direction--frequency grids and employed a U-Net architecture~\cite{ronneberger2015u}, whereas Hogg et
al.~\cite{hogg2024hrtf} combined a GAN~\cite{goodfellow2014generative} with a gnomonic equiangular projection.
Because HRTFs are inherently defined on a spherical spatial domain, planar
representations may introduce geometric distortion.
Alternative geometry-aware representations have therefore been explored, including
spherical CNNs~\cite{chen2023head,thuillier2024hrtf} and spherical-harmonic-domain modeling~\cite{hu2024hrtf,hu2025head}.

Despite their effectiveness, most learning-based methods formulate HRTF upsampling primarily as a spatial interpolation problem, focusing on relationships among sparsely sampled directions on the sphere.
In these methods, frequency-domain information is typically handled via independent per-frequency mappings or shallow aggregation, without explicitly modeling cross-frequency structure.
However, the log-magnitude spectrum exhibits rich frequency-dependent patterns, including pinna-induced resonances, spectral notches~\cite{algazi2001cipic}, and smooth cross-frequency correlations, which remain under-exploited.
Although frequency conditioning has been explored in recent work~\cite{ito2025spatial}, the contribution of different frequency-domain modeling strategies has not been systematically examined.

Motivated by this gap, we investigate frequency-domain feature modeling for
HRTF magnitude upsampling.
A frequency-domain modeling design space is considered, including per-frequency MLPs, convolutional architectures with increasing receptive fields, and attention-based models.
Based on this analysis, we propose frequency-domain Conformer (FD-Conformer), a sparse-to-dense HRTF upsampling model that combines convolution and self-attention~\cite{gulati2020conformer} to jointly model local spectral continuity and long-range frequency dependencies in HRTF magnitudes.
Experiments on the SONICOM~\cite{engel2023sonicom} and HUTUBS~\cite{brinkmann2019hutubs} datasets are carried out to demonstrate the superior performance of the proposed FD-Conformer over existing methods. 
The source code is available online~\cite{fdconformer_github2026}.

\begin{figure}[t] 
\centering 
\includegraphics[width=6cm]{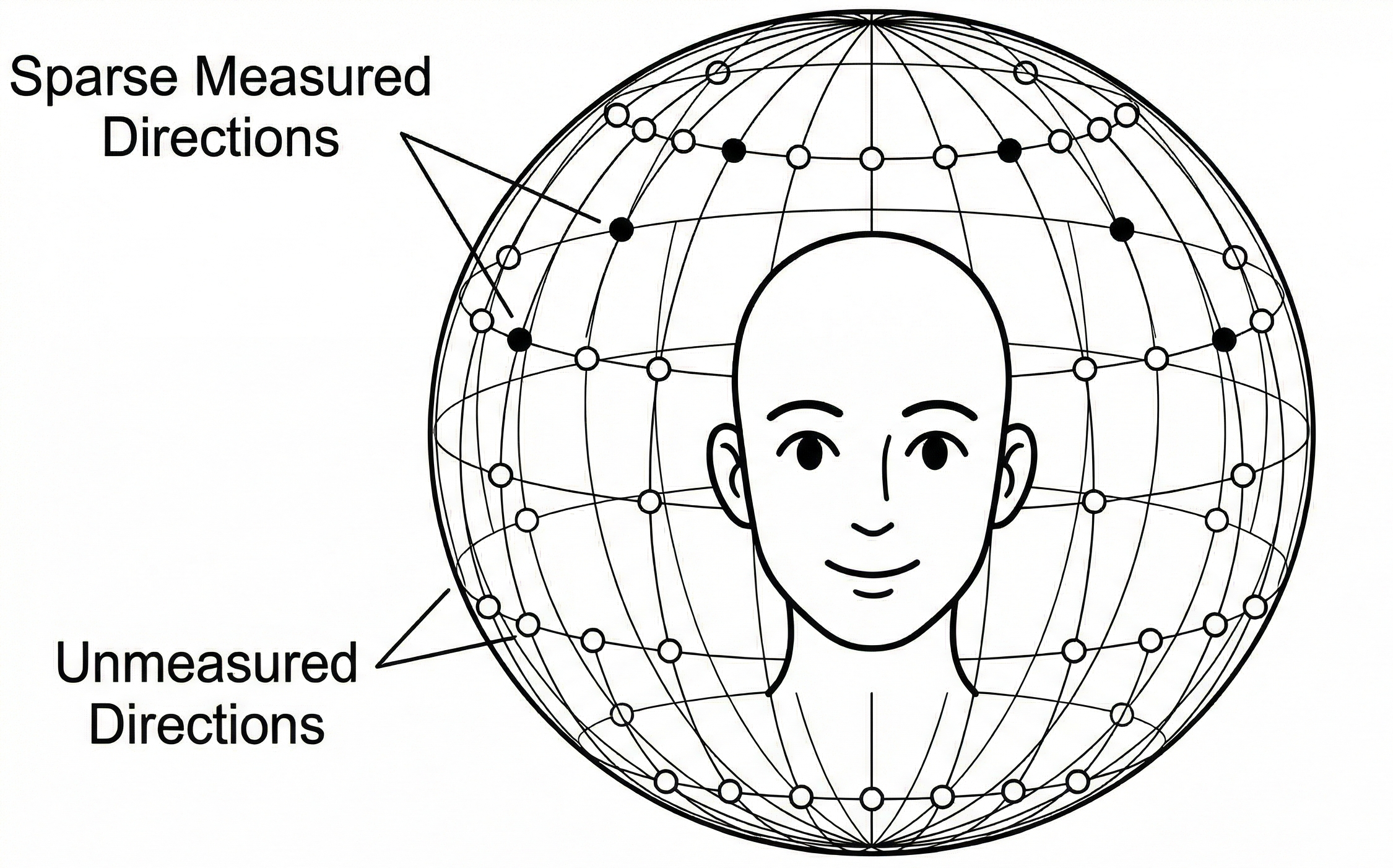} 
\caption{Conceptual illustration of HRTF upsampling from sparse measurements.} 
\label{fig:concept} 
\end{figure}

\section{Problem Statement}

Figure~\ref{fig:concept} illustrates the HRTF magnitude upsampling problem
considered in this work, where dense HRTFs are predicted from a limited number
of sparse measurements.
Let $\mathbf{d} \in \mathbb{S}^2$ denote a sound-source direction on the unit
sphere, parameterized by the azimuth $\theta \in [0, 2\pi)$ and elevation
$\phi \in [-\tfrac{\pi}{2}, \tfrac{\pi}{2}]$.
Let $\mathrm{H}(\mathbf{d},e,f) \in \mathbb{C}$ denote the complex-valued HRTF
from direction $\mathbf{d}$ to the left ($e=1$) or right ($e=2$) ear at
frequency $f$.
It is noted that we consider far-field HRTFs, for which the dependence on
source distance is negligible~\cite{brungart1999auditory} and therefore omitted.

To generate realistic spatial audio rendering effects, the HRTFs from a dense set $\mathcal{D}$ of source directions for a given subject are desired.
In practice, however, $\mathrm{H}(\mathbf{d},e,f)$ is available only at a sparse
subset $\mathcal{D}_M \subset \mathcal{D}$, where the cardinality
$|\mathcal{D}_M| = M \ll |\mathcal{D}|$ specifies the number of measured
directions.

Let $\mathcal{D}_U = \mathcal{D} \setminus \mathcal{D}_M$ denote the set of
unmeasured directions, with cardinality $|\mathcal{D}_U| = U$.
We focus on upsampling the log-magnitude spectrum of the HRTF, which is defined as
\begin{equation}
    \mathrm{H}_{\log}(\mathbf{d},e,f)
    = 20 \log_{10} \left| \mathrm{H}(\mathbf{d},e,f) \right|,
\end{equation}
where $|\cdot|$ denotes the magnitude operator, from the $M$ measured directions to the $U$ unmeasured directions.

Following the convention of learning-based methods for HRTF upsampling, we also assume access to a reference database containing densely measured $\mathrm{H}_{\log}$ from multiple subjects. 
These data provide patterns of how $\mathrm{H}_{\log}$ varies over direction and frequency across individuals, but only sparse measurements are available for the target subject.  
The goal is therefore to estimate $\hat{\mathrm{H}}_{\log}$ at the unmeasured directions $\mathcal{D}_U$ for the target subject, using only the sparse measurements and the patterns learned from the reference database.

\begin{figure}[t]
    \centering
    \includegraphics[width=0.7\linewidth]{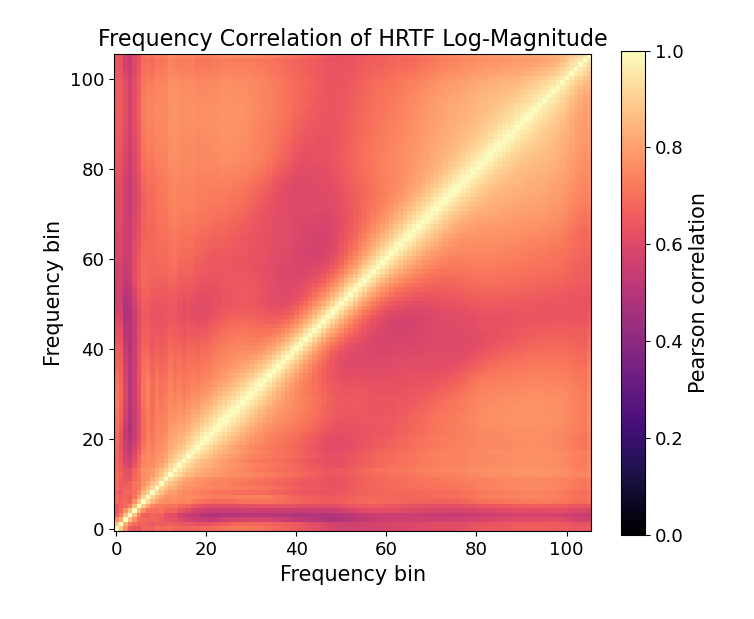}
    \caption{Frequency-frequency Pearson correlation of HRTF log-magnitude responses.
    Each entry indicates the correlation between two frequency bins, computed over
    all subjects, directions, and ears.
    Strong correlations are observed both locally around the diagonal and across
    distant frequency bins.}
    \label{fig:freq_corr}
\end{figure}

\begin{figure*}[t]
    \centering
    \includegraphics[width=\linewidth]{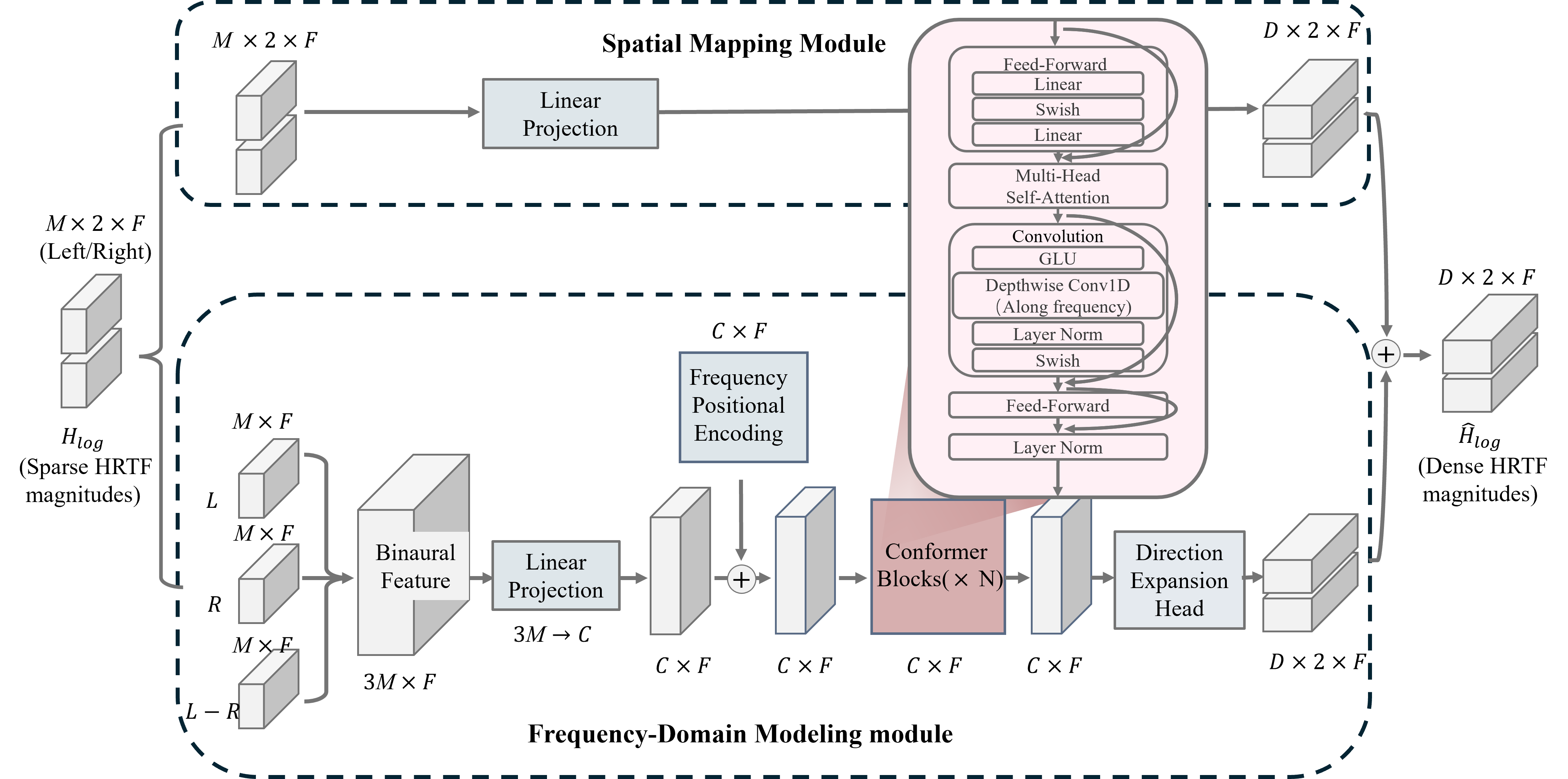}
    \caption{Overall framework of the proposed FD-Conformer. GLU denotes gated linear unit, and Conv1D denotes one-dimensional convolution.}
    \label{fig:model_architecture}
\end{figure*}

\section{Proposed Method}
\label{sec:method}

To motivate frequency-domain modeling, we first examine statistical dependencies
between frequency bins in HRTF log-magnitude spectra.
Figure~\ref{fig:freq_corr} shows the Pearson correlation matrix computed between
pairs of frequency bins using log-magnitude responses averaged over all subjects, directions, and ears based on the SONICOM dataset.
Strong spectral correlations are observed not only between neighboring frequencies but also across distant frequency bins, indicating the presence of both local spectral continuity and long-range frequency structure.

Motivated by this observation, we propose a frequency-domain neural architecture that explicitly models spectral dependencies along the frequency axis.
The overall framework of the proposed neural network (FD-Conformer) is illustrated in
Fig.~\ref{fig:model_architecture}.
Given sparse measurements $\mathrm{H}_{\log}(\mathbf{d},e,f)$ at directions $\mathbf{d} \in \mathcal{D}_M$, the network takes input as a tensor $\mathbf{X} \in \mathbb{R}^{M \times 2 \times F}$,
corresponding to $M$ measured directions, two ears, and $F$ frequency bins, and outputs the dense-direction HRTF log-magnitude tensor $\hat{\mathrm{H}}_{\log} \in \mathbb{R}^{D \times 2 \times F}$, including predictions at $U$ unmeasured directions.

\subsection{Network Architecture}

FD-Conformer employs a sparse-to-dense architecture comprising two modules: a spatial mapping module and a frequency-domain modeling module.
Both modules operate on the same sparse input and produce dense-direction estimates.
The final prediction is obtained by summing the outputs of the two modules:
\begin{equation}
\hat{\mathrm{H}}_{\log}
=
\hat{\mathrm{H}}_{\log}^{\mathrm{spatial}}
+
\hat{\mathrm{H}}_{\log}^{\mathrm{freq}} .
\label{eq:sum}
\end{equation}
The spatial mapping module focuses on direction-wise information aggregation across sparse measurements, whereas the frequency-domain modeling module captures spectral structure along the frequency dimension.
Combining the outputs of the two modules enables effective utilization of both spatial and spectral information.


The first component is the spatial mapping module, which performs a direct sparse-to-dense mapping across
directions using a linear transformation applied independently at each
frequency bin.
Given the input
$\mathbf{X} \in \mathbb{R}^{M \times 2 \times F}$, this module produces an intermediate dense estimate $\hat{\mathrm{H}}_{\log}^{\mathrm{spatial}} \in \mathbb{R}^{D \times 2 \times F}$.
This module captures direction-dependent structure by aggregating information
across sparse measurements, without introducing explicit modeling along the
frequency dimension.


\subsection{Frequency-Domain Modeling with Conformer Blocks}
\label{subsec:freq_modeling}

The second component is the frequency-domain modeling module.
It operates on a frequency-wise binaural
spectral representation derived from the sparse input.
For each frequency bin, three components are extracted from the sparse
measurements: the left-ear magnitude ($L$), the right-ear magnitude ($R$), and
their difference ($L - R$).
These components are concatenated along the feature dimension, yielding a
binaural spectral representation of dimension $3M$ per frequency bin.
Stacking the representations across all frequency bins results in a spectral
feature map $\mathbf{S} \in \mathbb{R}^{3M \times F}$, which preserves binaural
information while enabling explicit modeling along the frequency dimension.


The resulting binaural spectral representation $\mathbf{S}$ is first
projected into a latent spectral space of dimension $C$ and then processed
by a stack of Conformer blocks.
As self-attention is permutation-equivariant with respect to its input
sequence, it does not explicitly encode the frequency index or frequency-bin
ordering. We therefore add a learnable frequency positional encoding before the
Conformer blocks, following the learnable absolute position embeddings strategy commonly
used in transformer models~\cite{devlin2019bert}. This encoding is implemented
as a trainable tensor of size $F \times C$ and is added to the projected
spectral features, enabling the model to distinguish frequency bins and model
ordered spectral dependencies.

Let $\mathbf{H}^{(0)} \in \mathbb{R}^{F \times C}$ denote the resulting latent
spectral representation.
Frequency-domain modeling is performed using a stack of $N$ Conformer blocks,
which iteratively transform the features along the frequency axis:
\begin{equation}
\mathbf{H}^{(n)} =
\mathrm{ConformerBlock}\!\left(\mathbf{H}^{(n-1)}\right),
\quad n = 1, \ldots, N .
\end{equation}

Each Conformer block comprises two feed-forward networks (FFN), a multi-head
self-attention module (MHSA)~\cite{vaswani2017attention}, and a Conv, connected via
residual connections:
\begin{align}
\mathbf{H}_1 &= \mathbf{H}^{(n-1)} + \tfrac{1}{2}\,\mathrm{FFN}\!\left(\mathbf{H}^{(n-1)}\right), \\
\mathbf{H}_2 &= \mathbf{H}_1 + \mathrm{MHSA}\!\left(\mathbf{H}_1\right), \\
\mathbf{H}_3 &= \mathbf{H}_2 + \mathrm{Conv}\!\left(\mathbf{H}_2\right), \\
\mathbf{H}_4 &= \mathbf{H}_3 + \tfrac{1}{2}\,\mathrm{FFN}\!\left(\mathbf{H}_3\right), \\
\mathbf{H}^{(n)} &= \mathrm{LayerNorm}\!\left(\mathbf{H}_4\right).
\end{align}
In the proposed setting, $\mathrm{FFN}(\cdot)$ performs point-wise nonlinear
transformations in the latent spectral space at each frequency bin.
$\mathrm{MHSA}(\cdot)$ captures long-range dependencies across frequency bins,
which is useful for modeling global spectral structure in HRTF magnitudes.
$\mathrm{Conv}(\cdot)$ provides localized interactions along the frequency axis
and is well suited for representing local spectral continuity and fine spectral
details.
Together, these components enable joint modeling of local and long-range
frequency-domain features.

The output of the final Conformer block is mapped to dense-direction estimates through a direction expansion head, producing
$\hat{\mathrm{H}}_{\log}^{\mathrm{freq}} \in \mathbb{R}^{D \times 2 \times F}$.
The final output $\hat{\mathrm{H}}_{\log}$ is obtained by summing this term with
the output of the spatial mapping module, as defined in Eq.~\eqref{eq:sum}.

\subsection{Frequency-Domain Modeling Variants}
\label{subsec:freq_variants}
To systematically investigate the effect of frequency-domain modeling, we construct three model variants that share the same sparse-to-dense architecture as the proposed FD-Conformer. 
The variants differ only in the design of the frequency-domain modeling module, highlighted by the red blocks in Fig.~\ref{fig:model_architecture}.

The first variant employs a per-frequency multilayer perceptron (MLP), which
processes each frequency bin independently without modeling inter-frequency
dependencies. 
The second variant, denoted as \emph{Vanilla Conv}, employs a standard
one-dimensional convolutional module along the frequency axis to capture local
spectral interactions.
The third variant uses dilated convolution~\cite{yu2015multi} to enlarge the effective frequency receptive field and incorporate broader spectral context.

The proposed FD-Conformer combines convolution and self-attention to model both
local spectral interactions and long-range frequency dependencies. 
Together, these variants form a progressively more expressive frequency-domain design space, enabling a controlled comparison of different frequency-domain modeling strategies.

\subsection{Training Objective}

All model variants considered in this study are trained in a supervised manner.
Following common practice in HRTF magnitude upsampling, the log-spectral
distortion (LSD) is adopted as the primary training objective.
LSD measures the discrepancy between the predicted and ground-truth
log-magnitude spectra.
For each direction $\mathbf{d}$ and ear $e$, it is defined as
\begin{equation}
\mathrm{LSD}(\mathbf{d},e)
=
\sqrt{
\frac{1}{F}
\sum_{f=1}^{F}
\bigl(
\hat{\mathrm{H}}_{\log}(\mathbf{d},e,f)
-
\mathrm{H}_{\log}(\mathbf{d},e,f)
\bigr)^{2}
},
\label{eq:lsd_de}
\end{equation}
where $\mathrm{H}_{\log}(\mathbf{d},e,f)$ and
$\hat{\mathrm{H}}_{\log}(\mathbf{d},e,f)$ denote the ground-truth and predicted
log-magnitude responses at frequency bin $f$, respectively.
The overall LSD loss is obtained by averaging over both ears and all target
directions,
\begin{equation}
\mathcal{L}_{\mathrm{LSD}}
=
\frac{1}{2D}
\sum_{e=1}^{2}
\sum_{\mathbf{d} \in \mathcal{D}}
\mathrm{LSD}(\mathbf{d},e),
\end{equation}
where $\mathcal{D}$ denotes the set of dense directions.
The loss is computed independently for each subject and then averaged across
the training set.

Although LSD provides a global measure of log-magnitude reconstruction accuracy
averaged across frequencies, it treats individual frequency bins independently
and does not explicitly constrain local spectral variation.
As a result, sharp spectral features such as notches and peaks, which are
important for elevation perception and front--back discrimination, may not be
adequately preserved.

To complement the LSD objective, a spectral gradient loss (SGL) is introduced to
constrain first-order spectral variations along the frequency axis.
For each direction $\mathbf{d}$, ear $e$, and frequency bin $f$, the SGL is
defined as
\begin{equation}
\mathrm{SGL}(\mathbf{d},e,f)
=
\left|
\Delta_f \hat{\mathrm{H}}_{\log}(\mathbf{d},e,f)
-
\Delta_f \mathrm{H}_{\log}(\mathbf{d},e,f)
\right|,
\end{equation}
where
\begin{equation}
\Delta_f \mathrm{H}_{\log}(\mathbf{d},e,f)
=
\mathrm{H}_{\log}(\mathbf{d},e,f+1)
-
\mathrm{H}_{\log}(\mathbf{d},e,f)
\end{equation}
denotes the spectral difference between adjacent frequency bins.
The overall spectral gradient loss is obtained by averaging over directions,
ears, and frequencies, i.e.,
\begin{equation}
\mathcal{L}_{\mathrm{SGL}}
=
\frac{1}{2D(F-1)}
\sum_{e=1}^{2}
\sum_{\mathbf{d} \in \mathcal{D}}
\sum_{f=1}^{F-1}
\mathrm{SGL}(\mathbf{d},e,f).
\end{equation}

The total training objective is defined as
\begin{equation}
\mathcal{L}
=
\mathcal{L}_{\mathrm{LSD}}
+
\beta \mathcal{L}_{\mathrm{SGL}},
\end{equation}
where $\beta$ is a weighting factor balancing the two terms.
Preliminary experiments with different values of
$\beta$ showed that the proposed method is not sensitive to this weighting
factor; therefore, $\beta=1$ is adopted as a simple fixed setting without
further hyperparameter tuning.

\section{Experiments}

\subsection{Dataset and Preprocessing}

We evaluate the proposed method on the publicly available multi-subject HRTF dataset SONICOM~\cite{engel2023sonicom} first.  
The SONICOM dataset has since been extended to over $300$ subjects~\cite{poole2025extended}.
To maintain a consistent training and evaluation regime across all model variants and learning-based baselines, 
we use the first $200$ subjects following the listener acoustic personalization (LAP) Challenge protocol~\cite{hogg2025listener}, with the first $180$ subjects used for model development and the remaining $20$ for testing.
For these subjects, HRIR measurements are available at $D=793$ directions.
The HRIRs are recorded at 48~kHz and compensated using a minimum-phase free-field filter.
Following the LAP Challenge protocol~\cite{hogg2025listener}, each HRIR is transformed to the HRTF using a 256-point fast Fourier transform (FFT), and magnitudes within 187.5~Hz–19875~Hz are retained, yielding a 106-dimensional $\mathrm{H}_{\log}$ vector per ear.
Sparse-to-dense upsampling is evaluated under four sparse measurement configurations with $M \in \{3, 5, 19, 100\}$.

\subsection{Experimental Setup}

We use the Adam optimizer~\cite{kingma2014adam} with a learning rate of
$1\times10^{-3}$ and a batch size of $32$.
Training is performed for a maximum of $800$ epochs for all experiments.
The model checkpoint achieving the lowest validation LSD is selected for
evaluation.
During training, sparse measurements
$\mathrm{H}_{\log}(\mathbf{d},e,f)$ for $\mathbf{d} \in \mathcal{D}_M$ are provided
as input, and dense predictions are supervised using the full set of directions
$\mathcal{D}$.

The FD-Conformer employs $L=4$ Conformer blocks.
Each block uses a model dimension of $128$, a feed-forward dimension of $256$,
and $8$ attention heads.
The Conv in each Conformer block adopts a kernel size of $7$.
The direction expansion head consists of a two-layer multilayer perceptron with a
hidden dimension of $256$.
A dropout rate of $0.1$ is applied throughout the frequency-domain branch.
Learnable frequency positional encoding is enabled for all frequency-domain
modeling variants unless explicitly stated otherwise.


\subsection{Evaluation Metrics}

We evaluate the upsampling accuracy using two perceptually relevant metrics:
ILD error and LSD.
Both metrics are computed from the ground-truth log-magnitude spectra
$\mathrm{H}_{\log}(\mathbf{d},e,f)$ and the estimated spectra
$\hat{\mathrm{H}}_{\log}(\mathbf{d},e,f)$.

The ILD is evaluated as a broadband measure that captures the overall interaural
energy imbalance between the left- and right-ear responses.
We adopt a broadband ILD formulation that is consistent with the time-domain ILD
definition used in the LAP Challenge~\cite{hogg2025listener}.
The broadband energy at each ear is computed by summing the squared linear
magnitudes across all frequency bins:
\begin{equation}
E(\mathbf{d},e)
=
\sum_{f=1}^{F}
\left|
10^{\mathrm{H}_{\log}(\mathbf{d},e,f)/20}
\right|^{2}.
\end{equation}
The broadband ILD for direction $\mathbf{d}$ is then defined as the logarithmic
ratio between the left- and right-ear broadband energies:
\begin{equation}
\mathrm{ILD}(\mathbf{d})
=
10 \log_{10}
\frac{E(\mathbf{d},\text{Left})}{E(\mathbf{d},\text{Right})}.
\end{equation}
The estimated ILD, $\hat{\mathrm{ILD}}(\mathbf{d})$, is computed in the same
manner by replacing $E$ with the estimated broadband energy $\hat{E}$.

The ILD error is computed as the mean absolute difference between the
ground-truth and estimated ILD values over the unmeasured directions
$\mathcal{D}_U$:
\begin{equation}
\Delta \mathrm{ILD}
=
\frac{1}{U}
\sum_{\mathbf{d} \in \mathcal{D}_U}
\left|
\mathrm{ILD}(\mathbf{d}) - \hat{\mathrm{ILD}}(\mathbf{d})
\right|.
\end{equation}

The LSD is computed following Eq.~\eqref{eq:lsd_de} and averaged over both ears
and the unmeasured directions:
\begin{equation}
\mathrm{LSD}
=
\frac{1}{2U}
\sum_{e=1}^{2}
\sum_{\mathbf{d} \in \mathcal{D}_U}
\mathrm{LSD}(\mathbf{d},e).
\end{equation}

All metrics are computed exclusively on the unmeasured directions
$\mathcal{D}_U$, ensuring that the evaluation reflects upsampling performance rather than reconstruction of the measured inputs.

\begin{table*}[t]
\centering
\caption{Exploration of frequency-domain modeling strategies and ablation of the proposed model on the SONICOM dataset. Boldface values indicate the lowest errors for each metric and measurement setting.}
\label{tab:freq_design_space}
\resizebox{\textwidth}{!}{%
\begin{tabular}{l|cc|cc|cc|cc}
\hline
\multirow{2}{*}{Model Variant} &
\multicolumn{2}{c|}{3 measurements} &
\multicolumn{2}{c|}{5 measurements} &
\multicolumn{2}{c|}{19 measurements} &
\multicolumn{2}{c}{100 measurements} \\
& ILD [dB] & LSD [dB]
& ILD [dB] & LSD [dB]
& ILD [dB] & LSD [dB]
& ILD [dB] & LSD [dB] \\
\hline

Spatial mapping only
    & 2.42 & 5.83
    & 1.39 & 5.91
    & 0.88 & 3.59
    & 0.45 & 2.39 \\
\hline
\multicolumn{9}{l}{\textbf{Frequency-domain modeling strategies}} \\
\hline

Per-frequency MLP
    & 1.22 & 4.52
    & 1.15 & 4.91
    & 0.67 & 3.23
    & 0.41 & 2.27 \\

Vanilla Conv
    & 0.98 & 4.16
    & 1.15 & 4.51
    & \textbf{0.65} & 3.15
    & \textbf{0.40} & 2.27 \\

Dilated Conv (expanded receptive field)
    & 1.00 & 4.12
    & 1.02 & 4.34
    & 0.64 & 3.10
    & 0.42 & 2.27 \\
    
FD-Conformer (global $+$ local)
    & \textbf{0.94} & \textbf{3.97}
    & \textbf{0.97} & \textbf{4.00}
    & \textbf{0.65} & \textbf{3.08}
    & 0.41 & 2.27 \\
\hline

\multicolumn{9}{l}{\textbf{Conformer component ablations}} \\
\hline

w/o Conv
    & 1.12 & 4.59
    & 1.05 & 4.66
    & 0.67 & 3.17
    & 0.42 & 2.31 \\

w/o Positional Encoding
    & 0.96 & 4.08
    & 1.02 & 4.12
    & 0.66 & 3.09
    & 0.41 & 2.31 \\

w/o Spectral Gradient Loss
    & 0.95 & 3.97
    & 0.97 & 4.02
    & 0.66 & 3.08
    & 0.42 & 2.33 \\
\hline
\end{tabular}
}
\end{table*}

\subsection{Frequency-Domain Modeling Analysis}
\label{subsec:freq_modeling_analysis}

Table~\ref{tab:freq_design_space} reports the results of the frequency-domain
modeling variants introduced in Sec.~\ref{subsec:freq_variants}.
The upper part compares a spatial mapping only baseline, which relies solely on the spatial mapping module without any explicit frequency-domain modeling, with the three
alternative frequency-domain modeling variants and the proposed FD-Conformer, whereas the lower part presents ablations of its key components.

The spatial mapping only baseline consistently underperforms all
frequency-domain modeling variants.
In the extremely sparse settings (3 and 5 measurements), its LSD errors
are at least 1~dB higher than those with the frequency-domain modeling module, indicating the limitations of purely spatial
upsampling.

Among the frequency-domain modeling variants, the per-frequency MLP improves upon the
spatial-only baseline despite processing each frequency bin independently,
showing that frequency-wise nonlinear processing is beneficial even without
inter-frequency interactions. 
The convolutional variants further reduce LSD,
with \emph{Vanilla Conv} achieving 4.16~dB and 4.51~dB under 3 and 5 measurements, 
and dilated convolution further improves these values to
4.12~dB and 4.34~dB. 
This confirms the benefit of explicitly modeling local spectral dependencies and suggests that an expanded frequency receptive field is advantageous.

The FD-Conformer achieves the best overall performance, particularly under
extreme sparsity, with LSD errors of 3.97~dB and 4.00~dB for 3 and 5
measurements. 
Its advantage over convolutional variants suggests that combining local convolutional processing with long-range self-attention is more effective than relying solely on fixed convolutional kernels. 
As the number of sparse measurements increases, the performance gap between different frequency-domain modeling strategies gradually narrows. 
At 100 measurements, all methods achieve similar LSD values around 2.27~dB, implying that when sufficient directional information is available, the upsampling task becomes less sensitive to a
specific choice of frequency-domain modeling strategy.

The lower part of Table~\ref{tab:freq_design_space} presents an ablation study
of the FD-Conformer to assess the contribution of its key components.

Removing the Conv (\emph{w/o Conv}) leads to a clear degradation
in both LSD and ILD, particularly in the 3- and 5-measurement settings.
While the self-attention mechanism remains capable of modeling long-range
frequency dependencies, the absence of explicit local spectral interactions
significantly degrades performance under extreme sparsity.

Removing the learnable frequency positional encoding
(\emph{w/o Positional Encoding}) primarily affects low-sparsity conditions,
while having a negligible impact when more measurements are available.
This suggests that explicit frequency identity information becomes more relevant when spectral structure must be inferred from very limited directional input.
The relatively smaller degradation compared with removing the convolution module
may be because the Conv already introduces an ordering bias along the frequency axis through local neighborhood interactions.

Finally, removing the spectral gradient loss
(\emph{w/o Spectral Gradient Loss}) results in a modest performance decrease.
This indicates that while gradient-based regularization can help constrain local
spectral variation, the FD-Conformer architecture itself already provides strong
frequency-domain structure through its combined convolutional and
attention-based design.

Overall, these results highlight the importance of explicit frequency-domain
modeling in low-sparsity HRTF upsampling, whereas its relative impact diminishes as more directional measurements become available.

\begin{table*}[t]
\centering
\caption{SONICOM results: ILD and LSD errors for different numbers of sparse measurements. Boldface values indicate the lowest errors for each metric and measurement setting.}
\label{tab:main_results}
\resizebox{\textwidth}{!}{%
\begin{tabular}{l|cc|cc|cc|cc}
\hline
\multirow{2}{*}{Methods} &
\multicolumn{2}{c|}{3 measurements} &
\multicolumn{2}{c|}{5 measurements} &
\multicolumn{2}{c|}{19 measurements} &
\multicolumn{2}{c}{100 measurements} \\
& ILD [dB] & LSD [dB] 
& ILD [dB] & LSD [dB] 
& ILD [dB] & LSD [dB] 
& ILD [dB] & LSD [dB] \\
\hline
Barycentric        & 7.50 & 8.56 & 4.54 & 8.33 & 1.76 & 4.79 & 0.55 & 3.20 \\
Nearest neighbor        & 7.64 & 8.69 & 4.78 & 8.30 & 2.99 & 5.42 & 1.35 & 3.42 \\
SH  &  6.05 &  9.96 &  5.44 & 10.35 & 1.68 & 5.43 & 0.44& 3.38 \\
\hline

AE-GAN~\cite{hu2025head} & $1.38\pm 0.68$ & $4.84\pm 1.13$ & $1.32\pm 0.59$ & $4.76\pm 0.92$ & $0.98\pm 0.58$ & $4.28\pm 0.63$ &  $0.86\pm 0.29$ & $3.62\pm 0.39$ \\
IOA3D~\cite{zhao2025head} &  $1.18\pm 0.51$ &  $4.67\pm 0.42$ &  $1.28\pm 0.56$ &  $4.61\pm 0.54$  & $0.73\pm 0.25$ &  $3.21\pm 0.34$  & $0.56\pm 0.10$ &  $2.48\pm 0.22$ \\
RANF~\cite{masuyama2025retrieval} & $1.28 \pm 0.21$ & $4.53 \pm 0.39$& $1.31 \pm 0.26$ & $4.47 \pm 0.52$& $0.97 \pm 0.26$ & $3.65 \pm 0.42$& $0.78 \pm 0.11$ & $3.03 \pm 0.35$ \\
\hline
Proposed 
& $\boldsymbol{0.93 \pm 0.29}$ & $\boldsymbol{3.97 \pm 0.37}$
& $\boldsymbol{0.97 \pm 0.40}$ & $\boldsymbol{4.00 \pm 0.48}$
& $\boldsymbol{0.65 \pm 0.17}$ & $\boldsymbol{3.08 \pm 0.31}$
& $\boldsymbol{0.41 \pm 0.09}$ & $\boldsymbol{2.27 \pm 0.25}$
\\               
\hline
\end{tabular}
}
\end{table*}

\subsection{Comparative Evaluation with Existing Methods}
\label{subsec:eval_results}

Based on the architectural analysis presented in Sec.~\ref{subsec:freq_modeling_analysis}, we select the best-performing configuration of the proposed method, FD-Conformer, and compare it with existing approaches.
The evaluated baselines include traditional interpolation methods, such as
nearest-neighbor interpolation, barycentric interpolation, and spherical
harmonics (SH), as well as representative learning-based HRTF upsampling
methods. 
Specifically, RANF~\cite{masuyama2025retrieval} and
IOA3D~\cite{zhao2025head} are included because they ranked first and second,
respectively, in the LAP Challenge~\cite{hogg2025listener}. 
Autoencoder-based GAN (AE-GAN)~\cite{hu2025head} is included as a recent GAN-based method that performs HRTF upsampling in the SH domain and has demonstrated competitive performance through both objective evaluations and perceptual auditory model evaluations against traditional interpolation and other learning-based methods. 
Together, these methods represent highly competitive recent approaches to HRTF upsampling.
For a fair comparison, all learning-based baseline methods were retrained using the same training and test splits as the proposed method. 
No pretrained models or parameters from the original publications were used, thereby avoiding potential data leakage.

Table~\ref{tab:main_results} summarizes the quantitative results in terms of ILD error and LSD under different numbers of sparse measurements.

As additional references, we consider two non-personalized baselines.
First, a dataset-mean baseline is constructed by averaging the dense HRTFs across all training subjects and using the resulting mean HRTF as the prediction for each test subject.
This baseline gives an ILD error of 1.30~dB and an LSD of 4.90~dB on the SONICOM test set.

Second, we evaluate a random-subject baseline to account for the spectral smoothing introduced by population averaging. 
For each test subject, a complete dense HRTF is independently sampled from the training set and used as the prediction. 
Across 1000 random-assignment trials, this baseline yields an ILD
error of $1.77 \pm 0.12$~dB and an LSD of $6.81 \pm 0.14$~dB.
The substantially higher errors of both non-personalized baselines compared
with the learning-based methods suggest that the latter captures subject-specific
HRTF characteristics rather than relying on a generic population-level prior.

Traditional interpolation methods can provide accurate reconstruction when the spatial sampling density satisfies the spatial sampling theorem.
In the 100-measurement setting, their LSD and ILD errors are comparable to those
of learning-based approaches.
However, their performance rapidly degrades as the number of measurements
decreases, due to sampling limitations and their reliance on within-subject
interpolation only.

For the learning-based methods, the results are reported as the mean $\pm$ standard deviation across the test subjects.
Across all sparsity levels, FD-Conformer consistently achieves the lowest ILD and LSD errors among all compared methods.
In the extremely sparse setting with only 3 measurements, FD-Conformer achieves an LSD of $3.97\pm0.37$~dB, outperforming the strongest learning-based baselines, including RANF ($4.53\pm0.39$~dB), IOA3D ($4.67\pm0.42$~dB), and AE-GAN ($4.84\pm1.13$~dB), as well as all traditional interpolation methods.
With five measurements, FD-Conformer further reduces the LSD to $4.00$~dB, compared with $4.47$~dB for RANF, $4.61$~dB for IOA3D, and $4.76$~dB for AE-GAN.
As the number of available measurements increases, FD-Conformer continues to
deliver consistent improvements.
With 19 measured directions, it achieves an LSD of $3.08$\,dB, outperforming IOA3D ($3.21\pm0.34$~dB), RANF ($3.65\pm0.42$~dB), and AE-GAN ($4.28\pm0.63$~dB).
Even in the relatively dense setting with 100 measurements, FD-Conformer maintains the best performance, achieving an LSD of $2.27$\,dB.

The ILD results exhibit a similar trend.
Despite the smaller numerical range of ILD errors, FD-Conformer consistently outperforms all baseline methods across all sparsity levels,
indicating improved preservation of binaural level cues. Furthermore, paired Wilcoxon signed-rank tests based on the subject-level results indicate that the improvements over RANF are statistically significant for both ILD and LSD under all sparsity levels ($p<0.001$).

\begin{table}[t]
\centering
\caption{Computational complexity and inference time comparison. Params denotes the number of trainable parameters, while M and G represent millions ($10^6$) and billions ($10^9$), respectively.}
\label{tab:complexity}
\begin{tabular}{lccc}
\hline
Method & Params (M) & FLOPs (G) & Inference Time (ms) \\
\hline
AE-GAN~\cite{hu2025head} & 100.76 & 26.45 & 20.25\\
IOA3D~\cite{zhao2025head} & 4.63 & 1.96 & 0.39 \\
RANF~\cite{masuyama2025retrieval} & 1.23 & 110.79 & 50.06 \\
\hline
Proposed & 1.46 & 0.25 & 1.83 \\
\hline
\end{tabular}
\end{table}

In addition to upsampling accuracy, we compare the computational complexity and inference efficiency of the learning-based methods to assess their practical deployment cost.
Table~\ref{tab:complexity} summarizes the computational complexity and inference time of different learning-based methods under the 3 measurements setting. 
The inference time was measured on an NVIDIA L4 Tensor Core graphics processing unit (GPU; NVIDIA Corporation, Santa Clara, CA, USA) with a batch size of 1 and averaged over 500 forward passes after 100 warm-up iterations. 
For AE-GAN, only the generator is considered, because the discriminator is not used during inference.
For RANF, sparse measurements are first used to adapt the trained model to the target subject by fine-tuning a few subject-specific parameters. 
The reported inference time includes only the forward-pass cost after this adaptation and excludes the additional optimization required during the subject-adaptation stage. 
Therefore, the values in Table~\ref{tab:complexity} do not reflect the complete test-time computational cost of RANF.
The proposed method has the lowest number of floating-point operations (FLOPs) among the compared methods, requiring only 0.25 giga floating-point operations (GFLOPs). 
It uses 1.46 M parameters, which is much smaller than AE-GAN and IOA3D and comparable to RANF. 
Although the proposed model has lower FLOPs than IOA3D, its measured inference time is slightly higher. 
This difference stems from the distinct computational patterns of $\mathrm{MHSA}(\cdot)$ used in the proposed method and $\mathrm{Conv}(\cdot)$, which is mainly used in IOA3D, as convolution kernels are typically optimized more efficiently on modern GPU hardware. 
Overall, the proposed method achieves a favorable balance between model size, computational cost, and practical inference speed.

\begin{figure}[t]
    \centering

        \includegraphics[width=0.7\linewidth]{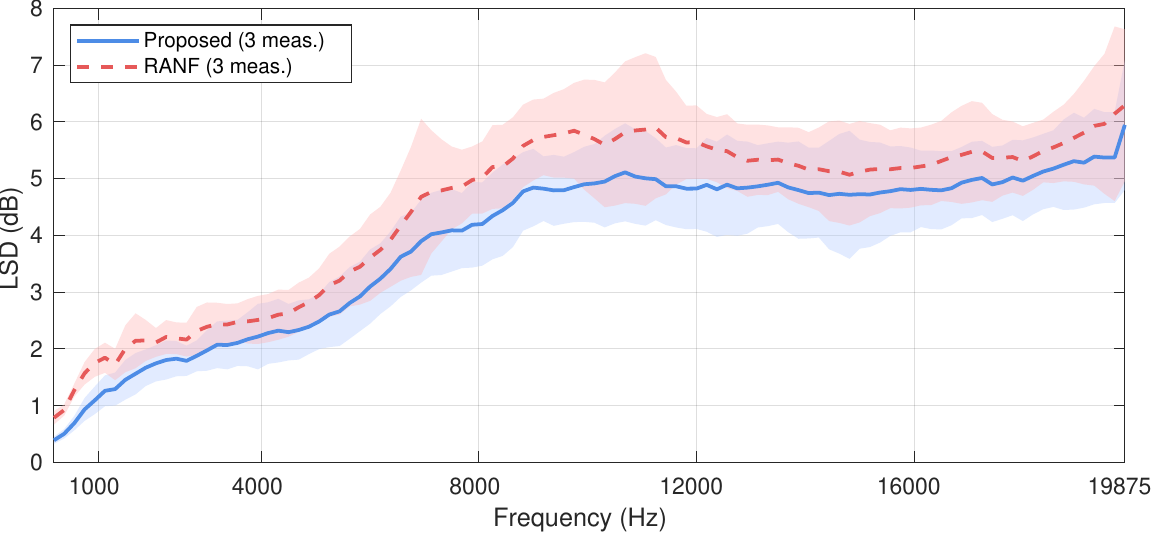} \\
        (a) 3 measurements 

        \includegraphics[width=0.7\linewidth]{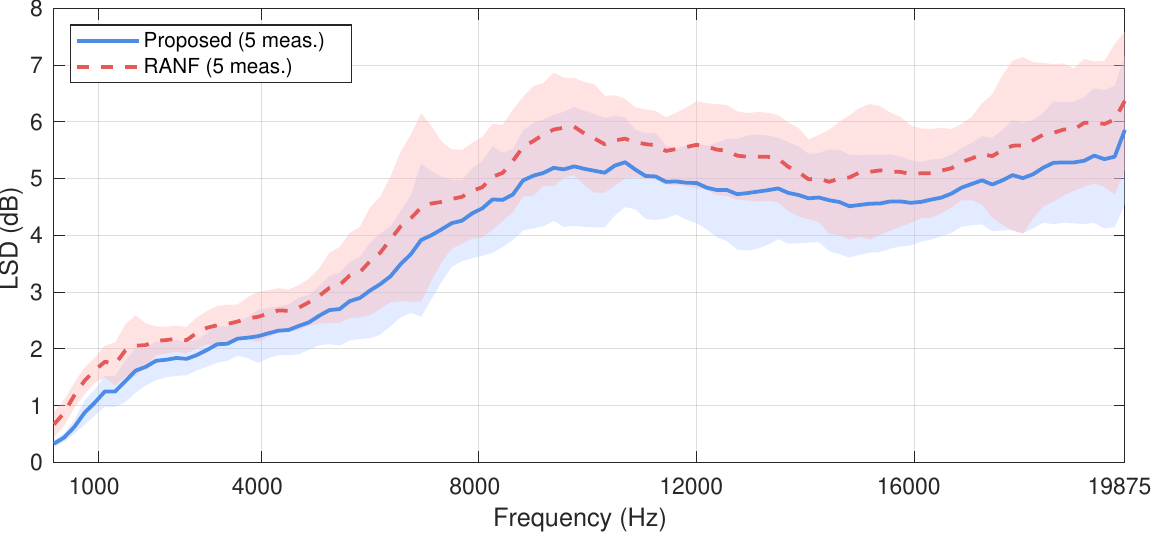} \\
        (b) 5 measurements 

        \includegraphics[width=0.7\linewidth]{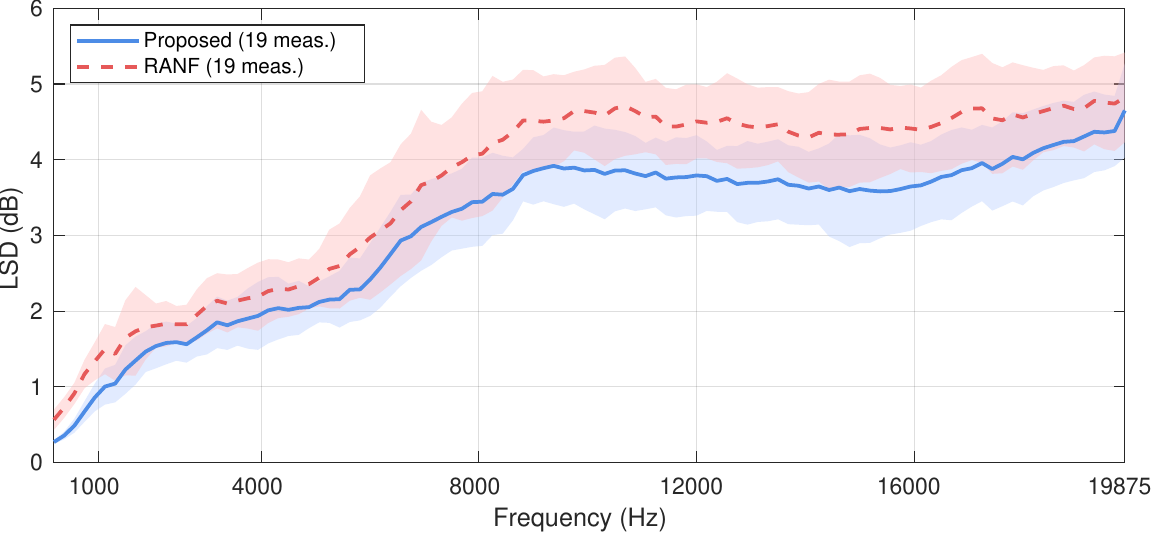} \\
        (c) 19 measurements

    \caption{Per-frequency LSD comparison between the proposed method and RANF under different numbers of measurements.
    The solid and dashed curves denote the mean LSD across the test subjects, whereas the shaded regions indicate one standard deviation. Measurements (meas.).}
    \label{fig:lsd}
\end{figure}

Figure~\ref{fig:lsd} presents the per-frequency LSD under different numbers of
sparse measurements.
The solid curves denote the mean LSD across the test subjects, while the shaded regions indicate one standard deviation.
Across most of the frequency range, FD-Conformer consistently yields lower LSD than RANF for all sparsity levels.
The performance gap is particularly pronounced in the mid- to high-frequency range (approximately 6--15~kHz), where HRTF magnitudes exhibit strong frequency-dependent structures such as pinna-induced resonances and spectral notches.
As the number of sparse measurements increases from 3 to 19, the LSD curves of both methods decrease overall; however, FD-Conformer maintains a clear advantage.

\begin{figure}[t]
    \centering
    \begin{tabular}{c}
        \includegraphics[width=\linewidth]{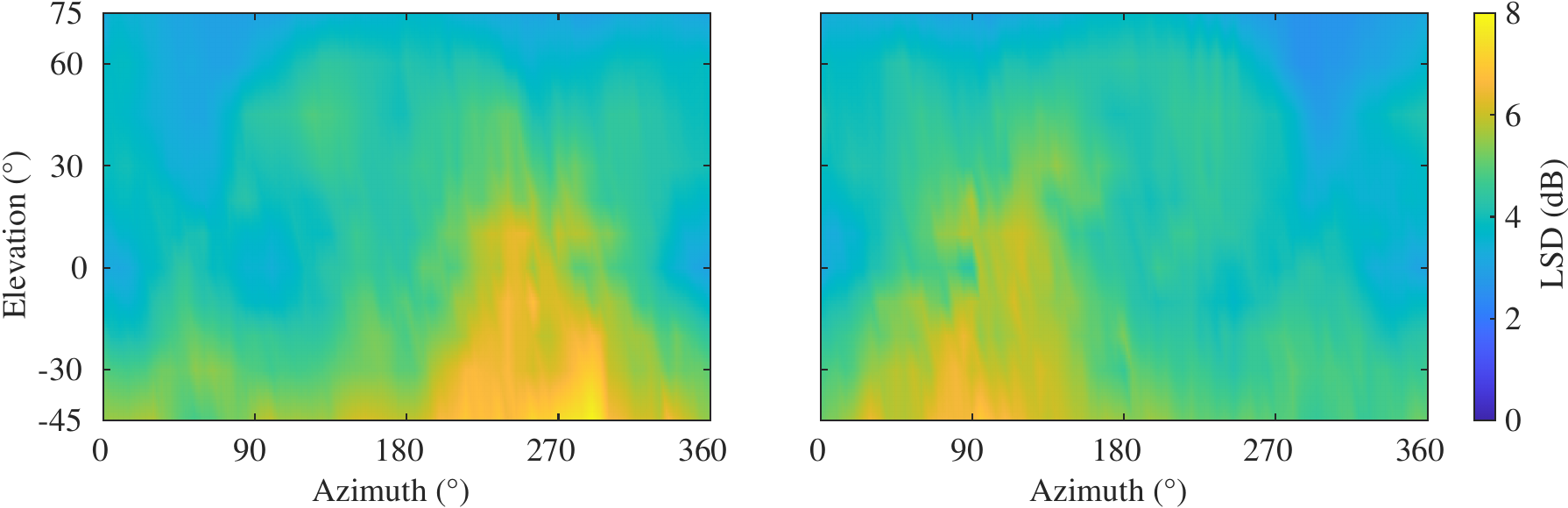} \\
        (a) RANF \\[0.8em]
        \includegraphics[width=\linewidth]{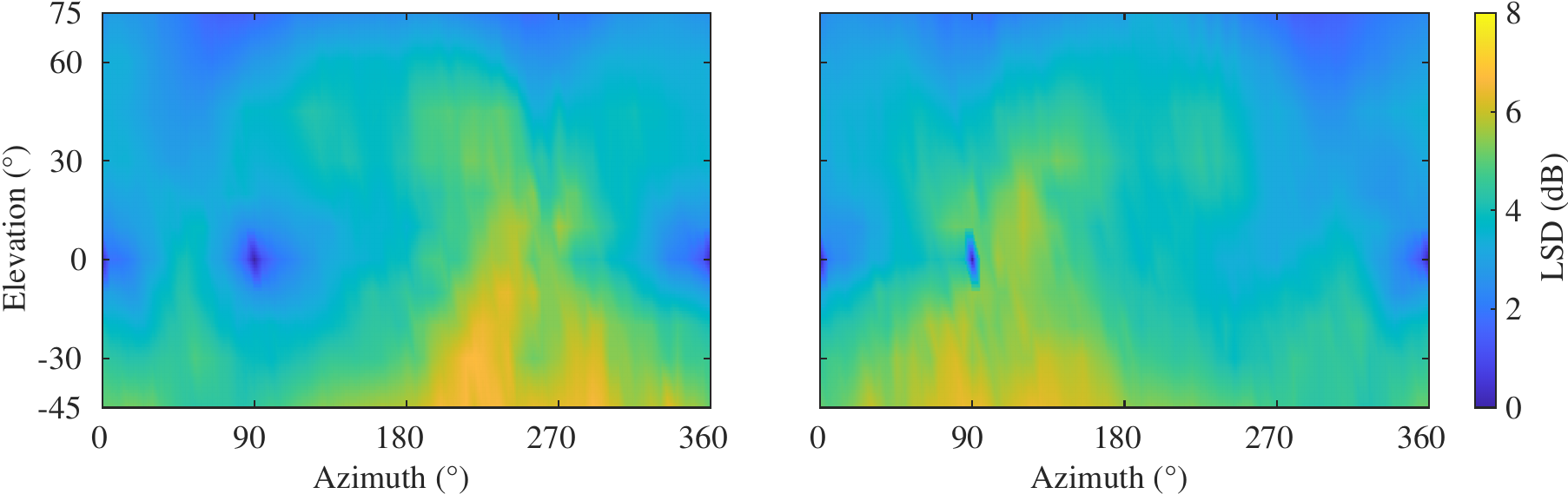} \\
        (b) FD-Conformer
    \end{tabular}
    \caption{Spatial distribution of LSD over azimuth
    and elevation under 3 measurements.
    For each method, the left and right panels correspond to the left and right
    ears, respectively.
    The visualization is limited to the highest
complete elevation ring at $\phi=75^\circ$. The single measurement direction
at $\phi=90^\circ$ is excluded to avoid representing it across azimuth.}
    \label{fig:lsd_spatial}
\end{figure}

Figure~\ref{fig:lsd_spatial} illustrates the spatial distribution of the LSD over azimuth and elevation under three sparse measurements, averaged over the whole frequency range. 
The visualization is limited to the highest complete elevation ring at $\phi=75^\circ$.
The azimuth follows the Spatially Oriented Format for Acoustics (SOFA) convention used in the dataset, where
$0^\circ$, $90^\circ$, $180^\circ$, and $270^\circ$ correspond to the frontal, left, rear, and right directions, respectively.
For both methods, larger reconstruction errors are consistently observed for contralateral source directions.
Compared with RANF, the proposed FD-Conformer produces lower LSD values across most directions for both ears.
The proportion of directions with LSD larger than 4~dB is reduced from
69.8\% for RANF to 45.6\% for FD-Conformer,
while the proportion with LSD larger than 5~dB is reduced from
27.5\% for RANF to 19.9\% for FD-Conformer.
In addition, the mean contralateral LSD is reduced from 4.90~dB
for RANF to 4.37~dB for FD-Conformer, supporting the
observation that regions with elevated reconstruction error are reduced.

\begin{figure*}[t]
\centering
\includegraphics[width=\linewidth]{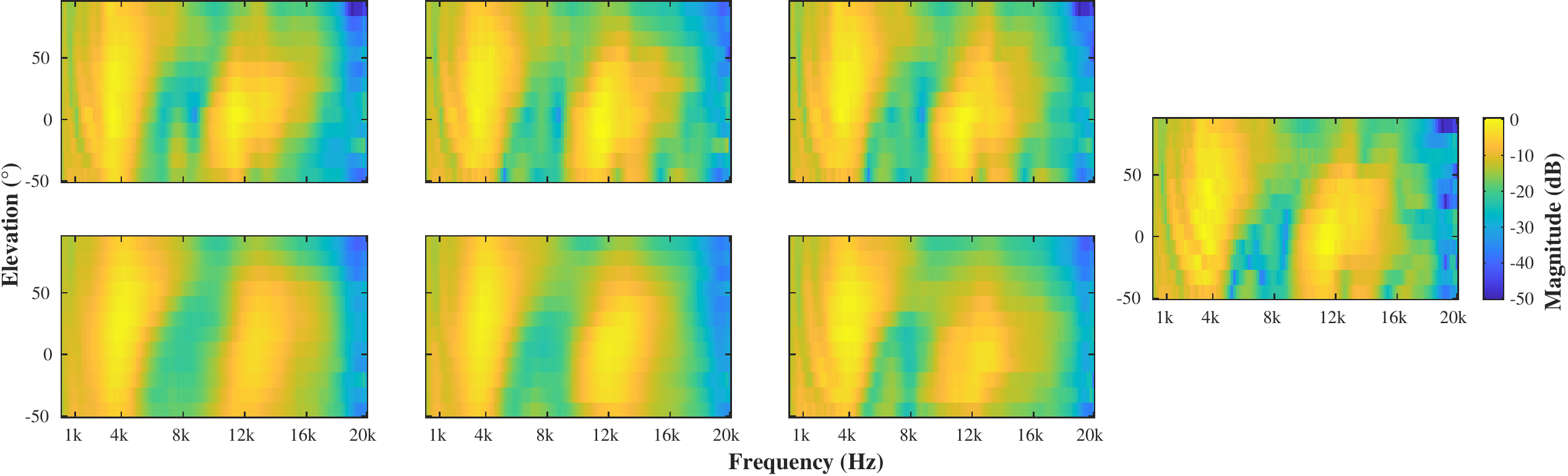}
\caption{
Median-plane elevation--frequency spectra for the left ear of subject \#200.
Top row: proposed FD-Conformer with 3, 5, and 19 sparse measurements.
Bottom row: RANF with the same sparse measurements.
Right: ground truth HRTF magnitude.
All panels share the same frequency and elevation axes as well as the same color scale.
For clarity, axis labels and ticks are shown only on the outermost panels, and the colorbar is shown only for the ground-truth panel.}

\label{fig:median_plane_layout}
\end{figure*}

Figure~\ref{fig:median_plane_layout} visualizes the median-plane
elevation--frequency spectra for a test subject (subject~\#200).
The top row shows the results of FD-Conformer with 3, 5, and 19 sparse
directions, while the bottom row presents the corresponding results obtained
using RANF.
The ground-truth spectrum is shown on the right for reference.
The reconstructed spectra exhibit qualitative differences between the two
methods, particularly in the mid- and high-frequency regions.
To quantify these differences, we computed the median-plane LSD over two frequency ranges: the mid-frequency band of 2--8~kHz and the high-frequency band of 8--20~kHz.
For the 3-measurement case, FD-Conformer
reduces the high-frequency LSD by 0.30~dB relative to RANF in the 8--20~kHz
band, although RANF gives a lower LSD in the 2--8~kHz band. With 5
measurements, FD-Conformer reduces the LSD relative to RANF by 0.60~dB in the
2--8~kHz band and by 0.62~dB in the 8--20~kHz band. With 19 measurements, the
corresponding reductions relative to RANF are 0.95~dB and 1.06~dB,
respectively.
These quantitative results indicate an overall advantage of FD-Conformer in
median-plane reconstruction, particularly under the 5- and 19-measurement
settings and in the high-frequency region.

\begin{table}[t]
\centering
\caption{HUTUBS results: ILD and LSD errors for different numbers of sparse measurements.}
\label{tab:hutubs_results}
\begin{tabular}{l|cc|cc}
\hline
\multirow{2}{*}{Methods} &
\multicolumn{2}{c|}{6 measurements} &
\multicolumn{2}{c}{14 measurements} \\
& ILD [dB] & LSD [dB]
& ILD [dB] & LSD [dB] \\
\hline
IOA3D~\cite{zhao2025head}              &  $1.00 \pm 0.08$ & $4.53 \pm 0.08$ & $0.80 \pm 0.05$ & $3.87 \pm 0.07$ \\
RANF~\cite{masuyama2025retrieval}             &  $1.22 \pm 0.31$ & $4.57 \pm 0.33$ & $1.04 \pm 0.13$ & $4.28 \pm 0.26$ \\

\hline
Proposed          & $0.79 \pm 0.07$ & $3.86 \pm 0.09$
                   & $0.67  \pm 0.05$ & $3.49 \pm 0.08$ \\
\hline
\end{tabular}
\end{table}

\subsection{HUTUBS Dataset}
\label{subsec:hutubs}

To further assess the robustness of the proposed method under different
measurement configurations, additional experiments are conducted on the
HUTUBS dataset~\cite{brinkmann2019hutubs}.
Compared to SONICOM, HUTUBS differs in both spatial sampling density and
measurement geometry, providing a complementary testbed for evaluating the
generalization of the proposed architecture.
The model is trained and evaluated separately on each dataset, and no
cross-dataset transfer is performed.

The HUTUBS dataset contains HRIR measurements from 96 subjects, each sampled at
440 spatial directions.
Following the same preprocessing pipeline as used for SONICOM, the HRIRs are
resampled to 32~kHz, zero-padded to 256 samples, and transformed to the
frequency domain using a 256-point FFT.
We retain 128 magnitude bins per ear, resulting in a 128-dimensional
$\mathrm{H}_{\log}$ representation.
After removing duplicate entries, 94 subjects remain, of which 77, 10, and 7
are used for training, validation, and testing, respectively.

To ensure consistency with the SONICOM experiments, the proposed method is
trained and evaluated on HUTUBS under the same experimental protocol.
Sparse measurement subsets are generated using downsampled Lebedev grids with
$M \in \{6, 14\}$.

As traditional interpolation-based methods exhibit trends similar to those
observed on SONICOM, Table~\ref{tab:hutubs_results} focuses on comparisons with the
strongest recent learning-based baselines, namely IOA3D and RANF.
AE-GAN is not included in the HUTUBS evaluation because its implementation is tied to the SONICOM sampling grid and spherical-harmonic configuration, and adapting it to HUTUBS would require architectural modifications beyond direct retraining.
For all learning-based methods, the results are reported as the mean
$\pm$ standard deviation across the test subjects.

As shown in Table~\ref{tab:hutubs_results}, the proposed method consistently
achieves the lowest ILD and LSD errors under both sparsity levels.
With 6 measurements, FD-Conformer reduces the LSD from
$4.53\pm0.08$~dB (IOA3D) and $4.57\pm0.33$~dB (RANF) to
$3.86\pm0.09$~dB, while simultaneously reducing the ILD error to
$0.79\pm0.07$~dB.
With 14 measurements, the proposed method further reduces the LSD to
$3.49\pm0.08$~dB and the ILD error to $0.67\pm0.05$~dB, outperforming both
IOA3D and RANF.
Furthermore, paired Wilcoxon signed-rank tests based on the subject-level
results indicate that the improvements over both baselines are statistically significant
for both ILD and LSD under both measurement settings ($p=0.0156$).
Overall, these results confirm that the benefits of explicit frequency-domain modeling extend beyond the SONICOM dataset, and the proposed method remains
robust under different spatial sampling patterns and dataset characteristics.

\section{Discussion and Limitations}
\label{sec:Discussion}

The results suggest that HRTF upsampling should not be regarded solely as
a spatial interpolation problem. 
Although most existing learning-based methods primarily emphasize spatial relationships across directions, our findings indicate that explicit frequency-domain modeling constitutes an important complementary design dimension, particularly under extremely sparse
measurement conditions where directional information is limited. 
LSD generally increases
toward higher frequencies, consistent with the increasing angular complexity of
HRTFs, where shorter acoustic wavelengths produce finer spatial variations and
sharper spectral features.
Under such sparse measurement conditions, spectral correlations provide a strong prior for upsampling. 
The ablation study further suggests that local spectral continuity and long-range frequency dependencies play complementary roles:
the Conv introduces an effective inductive bias for local
frequency structure, whereas self-attention captures broader spectral
correlations that cannot be represented by fixed receptive fields alone.
Although learning-based methods can exploit
population-level statistical priors to improve reconstruction beyond
conventional interpolation, their performance remains constrained by the
information available from sparse measurements.

The objective improvements observed in ILD and LSD may have meaningful
implications for auditory localization, although these implications remain
to be perceptually validated. More accurate ILD reconstruction may support
horizontal localization, while reduced mid- and high-frequency spectral
errors may better preserve pinna-related cues that contribute to elevation
perception and front--back discrimination. However, these conclusions are
inferred from objective metrics only, as no listening test or auditory
localization model is included in the present study. 
Although LSD is useful for comparing HRTF reconstruction performance at the
group level, recent work has reported little correlation between LSD and
behavioral localization accuracy at the individual level~\cite{pirard2026photogrammetry}.
Moreover, the current
formulation is limited to far-field log-magnitude upsampling and does not
explicitly evaluate phase, ITD estimation, or time-domain HRIR
reconstruction. 
Future perceptual experiments and binaural-signal evaluations
would help determine whether, and to what extent, the reported objective
gains translate into improved spatial-audio perception.

The practical generalization of the proposed method also remains to be
established under realistic and cross-dataset conditions. The experiments
are conducted on preprocessed public HRTF datasets, whereas real-world
measurements may be affected by microphone noise, positioning errors, room
reflections, incomplete compensation, and subject movement. Although the
consistent improvements on SONICOM and HUTUBS indicate that the proposed
frequency-domain modeling strategy is effective across different datasets
and measurement configurations, the model is trained and evaluated
separately within each dataset. Consequently, the present study does not
demonstrate cross-dataset generalization or robustness to noisy and
mismatched acquisition conditions. 
Previous studies have shown that differences between HRTF measurement setups
can affect the performance of learning-based models~\cite{pauwels2023relevance},
and recent work has investigated HRTF dataset harmonization and its perceptual
relevance~\cite{daugintis2025harmonization}.
Building on these studies, further work could investigate noise-aware training
and mixed-dataset evaluation for the proposed frequency-domain modeling
framework.

The current architecture is further constrained by its grid-dependent
formulation and late fusion design. The spatial mapping module operates on
fixed directional indices rather than explicit spatial coordinates, tying
the model to predefined sparse and dense measurement geometries. In
addition, the spatial and frequency-domain branches are combined only
through late additive fusion, which may limit the interaction between
spatial and spectral representations. As a result, the method is not
grid-agnostic and cannot be directly applied to arbitrary measurement
layouts without retraining or additional spatial alignment. Future work
could address these limitations through coordinate-conditioned or
geometry-aware modules, together with earlier fusion mechanisms such as
cross-attention or joint frequency--space representations.

\section{Conclusion}
\label{sec:conclusion}
This work investigated frequency-domain feature modeling for HRTF magnitude
upsampling from sparse directional measurements.
Although existing learning-based methods have primarily focused on modeling
spatial relationships across directions, the results demonstrate that the design of frequency-domain feature modeling plays an important role,
particularly under extremely sparse measurement conditions.
We proposed a sparse-to-dense framework that separates direct spatial mapping
from frequency-domain modeling.
Within this framework, a frequency-domain design space was explored, ranging
from per-frequency MLPs and convolutional models to a Conformer-based
architecture that jointly captures local spectral continuity and long-range
frequency dependencies.
Experiments on the SONICOM and HUTUBS datasets showed that progressively expanding the effective frequency receptive field leads to consistent performance improvements, with the Conformer-based model achieving the lowest ILD and LSD errors across all sparsity levels.
These findings underscore the importance of frequency-aware modeling
for HRTF magnitude upsampling and suggest that architectural choices along the frequency dimension should be regarded as a central design consideration,
rather than an implementation detail.
Future work will investigate tighter integration of frequency-domain modeling with spatial modeling, enabling joint learning of spectral and spatial structure.

\section*{Acknowledgments}

Computational facilities were provided by the University of Technology Sydney eResearch High Performance Computing Cluster.

\section*{AUTHOR DECLARATIONS}
\subsection*{Conflict of Interest}
The authors have no conflicts to disclose. 

\section*{DATA AVAILABILITY}
The data that support the findings of this study are publicly available at
\url{https://doi.org/10.5281/zenodo.21134348}. The HRTF datasets used in this
study are publicly available from the original sources cited in this article.


\bibliography{sampbib}

@inproceedings{wenzel1993perceptual,
  title = {Perceptual consequences of interpolating head-related transfer functions during spatial synthesis},
  author = {Wenzel, Elizabeth M and Foster, Scott H},
  booktitle = {Proceedings of the IEEE Workshop on Applications of Signal Processing to Audio and Acoustics},
  organization = {IEEE},
  pages = {102--105},
  address = {New Paltz, New York},
  month = {October 17--20},
  year = {1993},
}

@article{keyrouz2007binaural,
  title = {Binaural source localization and spatial audio reproduction for telepresence applications},
  author = {Keyrouz, Fakheredine and Diepold, Klaus},
  journal = {PRESENCE: Teleoperators and Virtual Environments},
  volume = {16},
  number = {5},
  pages = {509--522},
  year = {2007},
  publisher = {MIT Press One Rogers Street, Cambridge, MA 02142-1209, USA journals-info~…},
}

@article{geronazzo2018we,
  title = {Do we need individual head-related transfer functions for vertical localization? The case study of a spectral notch distance metric},
  author = {Geronazzo, Michele and Spagnol, Simone and Avanzini, Federico},
  journal = {IEEE/ACM Transactions on Audio, Speech, and Language Processing},
  volume = {26},
  number = {7},
  pages = {1247--1260},
  year = {2018},
  publisher = {IEEE},
}

@article{pelzer2020head,
  title = {Head-related transfer function recommendation based on perceptual similarities and anthropometric features},
  author = {Pelzer, Robert and Dinakaran, Manoj and Brinkmann, Fabian and Lepa, Steffen and Grosche, Peter and Weinzierl, Stefan},
  journal = {The Journal of the Acoustical Society of America},
  volume = {148},
  number = {6},
  pages = {3809--3817},
  year = {2020},
  publisher = {Acoustical Society of America},
}

@article{oberem2020experiments,
  title = {Experiments on localization accuracy with non-individual and individual {HRTFs} comparing static and dynamic reproduction methods},
  author = {Oberem, Josefa and Richter, Jan-Gerrit and Setzer, Dorothea and Seibold, Julia and Koch, Iring and Fels, Janina},
  journal = {bioRxiv:2020.03.31.011650},
  year = {2020},
  publisher = {Cold Spring Harbor Laboratory},
}

@article{andreopoulou2017identification,
  title = {Identification of perceptually relevant methods of inter-aural time difference estimation},
  author = {Andreopoulou, Areti and Katz, Brian FG},
  journal = {The Journal of the Acoustical Society of America},
  volume = {142},
  number = {2},
  pages = {588--598},
  year = {2017},
  publisher = {AIP Publishing},
}

@article{mckenzie2019interaural,
  title = {Interaural level difference optimization of binaural ambisonic rendering},
  author = {McKenzie, Thomas and Murphy, Damian T and Kearney, Gavin},
  journal = {Applied Sciences},
  volume = {9},
  number = {6},
  pages = {1226},
  year = {2019},
  publisher = {MDPI},
}

@article{carlini2024auditory,
  title = {Auditory localization: a comprehensive practical review},
  author = {Carlini, Alessandro and Bordeau, Camille and Ambard, Maxime},
  journal = {Frontiers in Psychology},
  volume = {15},
  pages = {1408073},
  year = {2024},
  publisher = {Frontiers Media SA},
}

@article{mcmullen2022machine,
  title = {A machine learning tutorial for spatial auditory display using head-related transfer functions},
  author = {McMullen, Kyla and Wan, Yunhao},
  journal = {The Journal of the Acoustical Society of America},
  volume = {151},
  number = {2},
  pages = {1277--1293},
  year = {2022},
  publisher = {Acoustical Society of America},
}

@article{fantini2025survey,
  title = {A survey on machine learning techniques for head-related transfer function individualization},
  author = {Fantini, Davide and Geronazzo, Michele and Avanzini, Federico and Ntalampiras, Stavros},
  journal = {IEEE Open Journal of Signal Processing},
  volume = {6},
  pages = {30--56},
  year = {2025},
  publisher = {IEEE},
}

@article{zhao2025head,
  title = {Head-Related Transfer Function Upsampling With Spatial Extrapolation Features},
  author = {Zhao, Jiale and Yao, Dingding and Li, Junfeng},
  journal = {IEEE/ACM Transactions on Audio, Speech, and Language Processing},
  volume = {33},
  pages = {1034--1048},
  year = {2025},
  publisher = {IEEE},
}

@article{pulkki1997virtual,
  title = {Virtual sound source positioning using vector base amplitude panning},
  author = {Pulkki, Ville},
  journal = {Journal of the audio engineering society},
  volume = {45},
  number = {6},
  pages = {456--466},
  year = {1997},
  publisher = {Audio Engineering Society},
}

@techreport{begault20003,
  title = {3-D sound for virtual reality and multimedia},
  author = {Begault, Durand R and Trejo, Leonard J},
  institution = {NASA Ames Research Center},
  number = {NASA/TM-2000-209606},
  address = {Moffett Field, CA},
  year = {2000},
}

@article{freeland2004interpositional,
  title = {Interpositional transfer function for {3D}-sound generation},
  author = {Freeland, F{\'a}bio P and Biscainho, Luiz WP and Diniz, Paulo SR},
  journal = {Journal of the Audio Engineering Society},
  volume = {52},
  number = {9},
  pages = {915--930},
  year = {2004},
  publisher = {Audio Engineering Society},
}

@inproceedings{hartung1999comparison,
  title = {Comparison of different methods for the interpolation of head-related transfer functions},
  author = {Hartung, K. and Braasch, J. and Sterbing, S. J.},
  booktitle = {Proceedings of the 16th International Audio Engineering Society Conference on Spatial Sound Reproduction},
  organization = {AES},
  pages = {319--329},
  address = {Rovaniemi, Finland},
  month = {April 10--12},
  year = {1999},
}

@article{cuevas20193d,
  title = {{3D} Tune-In Toolkit: An open-source library for real-time binaural spatialisation},
  author = {Cuevas-Rodr{\'\i}guez, Mar{\'\i}a and Picinali, Lorenzo and Gonz{\'a}lez-Toledo, Daniel and Garre, Carlos and de la Rubia-Cuestas, Ernesto and Molina-Tanco, Luis and Reyes-Lecuona, Arcadio},
  journal = {PloS one},
  volume = {14},
  number = {3},
  pages = {e0211899},
  year = {2019},
  publisher = {Public Library of Science San Francisco, CA USA},
}

@inproceedings{zotkin2009regularized,
  title = {Regularized {HRTF} fitting using spherical harmonics},
  author = {Zotkin, D. N. and Duraiswami, R. and Gumerov, N. A.},
  booktitle = {Proceedings of IEEE Workshop on Applications of Signal Processing to Audio and Acoustics},
  organization = {IEEE},
  pages = {257--260},
  address = {New Paltz, New York},
  month = {October 18--21},
  year = {2009},
}

@inproceedings{ahrens2012hrtf,
  title = {{HRTF} magnitude modeling using a non-regularized least-squares fit of spherical harmonics coefficients on incomplete data},
  author = {Ahrens, J. and Thomas, M. R. P. and Tashev, I.},
  booktitle = {Proceedings of the 2012 Asia Pacific Signal and Information Processing Association Annual Summit and Conference},
  organization = {IEEE},
  pages = {1--5},
  address = {Hollywood, CA},
  month = {December 3--6},
  year = {2012},
}

@article{porschmann2019directional,
  title = {Directional equalization of sparse head-related transfer function sets for spatial upsampling},
  author = {P{\"o}rschmann, Christoph and Arend, Johannes M and Brinkmann, Fabian},
  journal = {IEEE/ACM Transactions on Audio, Speech, and Language Processing},
  volume = {27},
  number = {6},
  pages = {1060--1071},
  year = {2019},
  publisher = {IEEE},
}

@article{arend2021assessing,
  title = {Assessing spherical harmonics interpolation of time-aligned head-related transfer functions},
  author = {Arend, Johannes M and Brinkmann, Fabian and P{\"o}rschmann, Christoph},
  journal = {Journal of the Audio Engineering Society},
  volume = {69},
  number = {1/2},
  pages = {104--117},
  year = {2021},
  publisher = {Audio Engineering Society},
}

@article{xie2012recovery,
  title = {Recovery of individual head-related transfer functions from a small set of measurements},
  author = {Xie, B.-S.},
  journal = {J. Acoust. Soc. Amer.},
  volume = {132},
  number = {1},
  pages = {282--294},
  year = {2012},
  publisher = {AIP Publishing},
}

@article{zhang2020modeling,
  title = {Modeling of individual {HRTFs} based on spatial principal component analysis},
  author = {Zhang, M. and Ge, Z. and Liu, T. and Wu, X. and Qu, T.},
  journal = {IEEE/ACM Trans. Audio, Speech, Language Process.},
  volume = {28},
  pages = {785--797},
  year = {2020},
  publisher = {IEEE},
}

@article{watanabe2014dataset,
  title = {Dataset of head-related transfer functions measured with a circular loudspeaker array},
  author = {Watanabe, Kanji and Iwaya, Yukio and Suzuki, Y{\^o}iti and Takane, Shouichi and Sato, Sojun},
  journal = {Acoustical science and technology},
  volume = {35},
  number = {3},
  pages = {159--165},
  year = {2014},
  publisher = {Acoustical Society of Japan},
}

@inproceedings{sridhar2017database,
  title = {A database of head-related transfer functions and morphological measurements},
  author = {Sridhar, Rahulram and Tylka, Joseph G and Choueiri, Edgar},
  booktitle = {Proceedings of the Audio Engineering Society Convention 143},
  organization = {AES},
  pages = {851--855},
  address = {New York, New York},
  month = {October 18--21},
  year = {2017},
}

@article{armstrong2018perceptual,
  title = {A perceptual evaluation of individual and non-individual {HRTFs}: A case study of the {SADIE} II database},
  author = {Armstrong, Cal and Thresh, Lewis and Murphy, Damian and Kearney, Gavin},
  journal = {Applied Sciences},
  volume = {8},
  number = {11},
  pages = {2029},
  year = {2018},
  publisher = {MDPI},
}

@misc{brinkmann2019hutubs,
  title = {The {HUTUBS} {HRTF} database},
  author = {Brinkmann, F. and Dinakaran, M. and Pelzer, R. and Wohlgemuth, J. J. and Seipel, F. and Voss, D. and Grosche, P. and Weinzierl, S.},
  year = {2019},
  howpublished = {HRTF},
}

@article{engel2023sonicom,
  title = {The {SONICOM} {HRTF} dataset},
  author = {Engel, Isaac and Daugintis, Rapolas and Vicente, Thibault and Hogg, Aidan OT and Pauwels, Johan and Tournier, Arnaud J and Picinali, Lorenzo},
  journal = {journal of the audio engineering society},
  volume = {71},
  number = {5},
  pages = {241--253},
  year = {2023},
  publisher = {Audio Engineering Society},
}

@inproceedings{ito2022head,
  title = {Head-related transfer function interpolation from spatially sparse measurements using autoencoder with source position conditioning},
  author = {Ito, Yuki and Nakamura, Tomohiko and Koyama, Shoichi and Saruwatari, Hiroshi},
  booktitle = {Proceedings of the 2022 International Workshop on Acoustic Signal Enhancement (IWAENC)},
  organization = {IEEE},
  pages = {1--5},
  address = {Bamberg, Germany},
  month = {September 5--8},
  year = {2022},
}

@article{ito2025spatial,
  title = {Spatial upsampling of head-related transfer function using neural network conditioned on source position and frequency},
  author = {Ito, Yuki and Nakamura, Tomohiko and Koyama, Shoichi and Sakamoto, Shuichi and Saruwatari, Hiroshi},
  journal = {IEEE Open Journal of Signal Processing},
  volume = {6},
  pages = {1109--1123},
  year = {2025},
  publisher = {IEEE},
}

@inproceedings{zhang2023hrtf,
  title = {{HRTF} Field: Unifying Measured {HRTF} Magnitude Representation with Neural Fields},
  author = {Zhang, Y. and Wang, Y. and Duan, Z.},
  booktitle = {Proceedings of 2023 IEEE International Conference on Acoustics, Speech and Signal Processing (ICASSP)},
  organization = {IEEE},
  pages = {1--5},
  address = {Rhodes Island, Greece},
  month = {June 4--10},
  year = {2023},
}

@inproceedings{lee2023global,
  title = {Global {HRTF} interpolation via learned affine transformation of hyper-conditioned features},
  author = {Lee, Jin Woo and Lee, Sungho and Lee, Kyogu},
  booktitle = {Proceedings of 2023 IEEE International Conference on Acoustics, Speech and Signal Processing (ICASSP)},
  organization = {IEEE},
  pages = {1--5},
  address = {Rhodes Island, Greece},
  month = {June 4--10},
  year = {2023},
}

@article{masuyama2024niirf,
  title = {{NIIRF}: Neural {IIR} Filter Field for {HRTF} Upsampling and Personalization},
  author = {Masuyama, Yoshiki and Wichern, Gordon and Germain, Fran{\c{c}}ois G and Pan, Zexu and Khurana, Sameer and Hori, Chiori and Roux, Jonathan Le},
  journal = {arXiv preprint arXiv:2402.17907},
  year = {2024},
}

@article{mildenhall2021nerf,
  title = {{NeRF}: Representing scenes as neural radiance fields for view synthesis},
  author = {Mildenhall, Ben and Srinivasan, Pratul P and Tancik, Matthew and Barron, Jonathan T and Ramamoorthi, Ravi and Ng, Ren},
  journal = {Communications of the ACM},
  volume = {65},
  number = {1},
  pages = {99--106},
  year = {2022},
  publisher = {ACM New York, NY, USA},
}

@article{hu2025graph,
  title = {Graph Neural Field with Spatial-Correlation Augmentation for {HRTF} Personalization},
  author = {Hu, De and Hu, Junsheng and Jiang, Cuicui},
  journal = {arXiv preprint arXiv:2511.10697},
  year = {2025},
}

@inproceedings{masuyama2025retrieval,
  title = {Retrieval-Augmented Neural Field for {HRTF} Upsampling and Personalization},
  author = {Masuyama, Yoshiki and Wichern, Gordon and Germain, Fran{\c{c}}ois G and Ick, Christopher and Le Roux, Jonathan},
  booktitle = {Proceedings of the 2025 IEEE International Conference on Acoustics, Speech and Signal Processing (ICASSP)},
  organization = {IEEE},
  pages = {1--5},
  address = {Hyderabad, India},
  month = {April 6--11},
  year = {2025},
}

@article{ma2023spatial,
  title = {Spatial upsampling of head-related transfer functions using a physics-informed neural network},
  author = {Ma, Fei and Abhayapala, Thushara D and Samarasinghe, Prasanga N and Chen, Xingyu},
  journal = {arXiv preprint arXiv:2307.14650},
  year = {2023},
}

@inproceedings{zurale2023spatial,
  title = {Spatial upsampling of sparse head related transfer functions-a {VQ-VAE} \& transformer based approach},
  author = {Zurale, Devansh and Dubnov, Shlomo},
  booktitle = {Proceedings of the AES 2023 International Conference on Spatial and Immersive Audio},
  organization = {AES},
  address = {Huddersfield, UK},
  month = {August 23--25},
  year = {2023},
}

@article{hogg2024hrtf,
  title = {{HRTF} upsampling with a generative adversarial network using a gnomonic equiangular projection},
  author = {Hogg, Aidan OT and Jenkins, Mads and Liu, He and Squires, Isaac and Cooper, Samuel J and Picinali, Lorenzo},
  journal = {IEEE/ACM Transactions on Audio, Speech, and Language Processing},
  volume = {32},
  pages = {2085--2099},
  year = {2024},
  publisher = {IEEE},
}

@article{jiang2023modeling,
  title = {Modeling individual head-related transfer functions from sparse measurements using a convolutional neural network},
  author = {Jiang, Z. and Sang, J. and Zheng, C. and Li, A. and Li, X.},
  journal = {J. Acoust. Soc. Amer.},
  volume = {153},
  number = {1},
  pages = {248--259},
  year = {2023},
  publisher = {AIP Publishing},
}

@inproceedings{ronneberger2015u,
  title = {{U-Net}: Convolutional networks for biomedical image segmentation},
  author = {Ronneberger, Olaf and Fischer, Philipp and Brox, Thomas},
  booktitle = {Proceedings of the Medical Image Computing and Computer-Assisted Intervention 18th International Conference},
  organization = {Springer},
  pages = {234--241},
  address = {Munich, Germany},
  month = {October 5--9},
  year = {2015},
}

@article{goodfellow2014generative,
  title = {Generative adversarial nets},
  author = {Goodfellow, Ian J and Pouget-Abadie, Jean and Mirza, Mehdi and Xu, Bing and Warde-Farley, David and Ozair, Sherjil and Courville, Aaron and Bengio, Yoshua},
  journal = {Advances in Neural Information Processing Systems},
  volume = {27},
  pages = {2672--2680},
  year = {2014},
}

@article{chen2023head,
  title = {Head-related transfer function interpolation with a spherical {CNN}},
  author = {Chen, Xingyu and Ma, Fei and Zhang, Yile and Bastine, Amy and Samarasinghe, Prasanga N},
  journal = {arXiv preprint arXiv:2309.08290},
  year = {2023},
}

@article{thuillier2024hrtf,
  title = {{HRTF} Interpolation using a Spherical Neural Process Meta-Learner},
  author = {Thuillier, Etienne and Jin, Craig and V{\"a}lim{\"a}ki, Vesa},
  journal = {IEEE/ACM Transactions on Audio, Speech, and Language Processing},
  volume = {32},
  pages = {1790--1802},
  year = {2024},
  publisher = {IEEE},
}

@inproceedings{hu2024hrtf,
  title = {{HRTF} spatial upsampling in the spherical harmonics domain employing a generative adversarial network},
  author = {Hu, Xuyi and Li, Jian and Picinali, Lorenzo and Hogg, Aidan OT},
  booktitle = {Proceedings of the 27th International Conference on Digital Audio Effects (DAFx24)},
  pages = {396--403},
  address = {Guildford, Surrey, UK},
  month = {September 3--7},
  year = {2024},
}

@article{hu2025head,
  title = {Head-related transfer function upsampling using an autoencoder-based generative adversarial network with evaluation framework},
  author = {Hu, Xuyi and Li, Jian and Picinali, Lorenzo and Hogg, Aidan OT},
  journal = {Journal of the Audio Engineering Society},
  volume = {73},
  number = {9},
  pages = {533--547},
  year = {2025},
}

@inproceedings{algazi2001cipic,
  title = {The {CIPIC} {HRTF} database},
  author = {Algazi, V Ralph and Duda, Richard O and Thompson, Dennis M and Avendano, Carlos},
  booktitle = {Proceedings of the 2001 IEEE Workshop on the Applications of Signal Processing to Audio and Acoustics},
  organization = {IEEE},
  pages = {99--102},
  address = {New Paltz, New York},
  month = {October 21--24},
  year = {2001},
}

@article{gulati2020conformer,
  title = {Conformer: Convolution-augmented transformer for speech recognition},
  author = {Gulati, Anmol and Qin, James and Chiu, Chung-Cheng and Parmar, Niki and Zhang, Yu and Yu, Jiahui and Han, Wei and Wang, Shibo and Zhang, Zhengdong and Wu, Yonghui and Pang, Ruoming},
  journal = {arXiv preprint arXiv:2005.08100},
  year = {2020},
}

@misc{fdconformer_github2026,
  year = {2026},
  howpublished = {\url{https://github.com/xingyuaudio/FD-Conformer}},
  note = {Last viewed 21 October 2026},
}

@article{brungart1999auditory,
  title = {Auditory localization of nearby sources. Head-related transfer functions},
  author = {Brungart, Douglas S and Rabinowitz, William M},
  journal = {The Journal of the Acoustical Society of America},
  volume = {106},
  number = {3},
  pages = {1465--1479},
  year = {1999},
  publisher = {Acoustical Society of America},
}

@inproceedings{devlin2019bert,
  title = {{BERT}: Pre-training of deep bidirectional transformers for language understanding},
  author = {Devlin, Jacob and Chang, Ming-Wei and Lee, Kenton and Toutanova, Kristina},
  booktitle = {Proceedings of the 2019 Conference of the North American Chapter of the Association for Computational Linguistics: Human Language Technologies, Volume 1 (Long and Short Papers)},
  organization = {Association for Computational Linguistics (ACL)},
  pages = {4171--4186},
  address = {Minneapolis, MN},
  month = {June 2--7},
  year = {2019},
}

@article{vaswani2017attention,
  title = {Attention is all you need},
  author = {Vaswani, Ashish and Shazeer, Noam and Parmar, Niki and Uszkoreit, Jakob and Jones, Llion and Gomez, Aidan N and Kaiser, {\L}ukasz and Polosukhin, Illia},
  journal = {Advances in Neural Information Processing Systems},
  volume = {30},
  pages = {5998--6008},
  year = {2017},
}

@article{yu2015multi,
  title = {Multi-scale context aggregation by dilated convolutions},
  author = {Yu, Fisher and Koltun, Vladlen},
  journal = {arXiv preprint arXiv:1511.07122},
  year = {2015},
}

@inproceedings{poole2025extended,
  title = {The extended {SONICOM} {HRTF} dataset and spatial audio metrics toolbox},
  author = {Poole, Katarina C. and Meyer, Julie and Martin, Vincent and Daugintis, Rapolas and Marggraf-Turley, Nils and Webb, Jack and Pirard, Ludovic and La Magna, Nicola and Picinali, Lorenzo},
  booktitle = {Proceedings of the 11th Convention of the European Acoustics Association: Forum Acusticum},
  organization = {EAA},
  pages = {6483--6486},
  address = {M\'{a}laga, Spain},
  month = {June 23--26},
  year = {2025},
}

@article{hogg2025listener,
  title = {Listener acoustic personalisation challenge-{LAP24}: Head-related transfer function upsampling},
  author = {Hogg, Aidan OT and Barumerli, Roberto and Daugintis, Rapolas and Poole, Katarina C and Brinkmann, Fabian and Picinali, Lorenzo and Geronazzo, Michele},
  journal = {IEEE Open Journal of Signal Processing},
  volume = {6},
  pages = {926--941},
  year = {2025},
  publisher = {IEEE},
}

@inproceedings{kingma2014adam,
  title = {Adam: A Method for Stochastic Optimization},
  author = {Kingma, Diederik P and Ba, Jimmy},
  booktitle = {Proceedings of the International Conference on Learning Representations (ICLR)},
  pages = {1--15},
  address = {San Diego, CA},
  month = {May 7--9},
  year = {2015},
}

@misc{pirard2026photogrammetry,
  title = {Photogrammetry-Reconstructed {3D} Head Meshes for Accessible Individual Head-Related Transfer Functions},
  author = {Pirard, Ludovic and Picinali, Lorenzo and Poole, Katarina C.},
  year = {2026},
  howpublished = {arXiv:2603.24104},
}

@inproceedings{pauwels2023relevance,
  title = {On the relevance of the differences between {HRTF} measurement setups for machine learning},
  author = {Pauwels, J. and Picinali, L.},
  booktitle = {Proceedings of the 2023 IEEE International Conference on Acoustics, Speech and Signal Processing (ICASSP)},
  organization = {IEEE},
  pages = {1--5},
  address = {Rhodes Island, Greece},
  month = {June 4--10},
  year = {2023},
}

@article{daugintis2025harmonization,
  title = {Listener Acoustic Personalization Challenge---{LAP24}: Head-Related Transfer Function Dataset Harmonization},
  author = {Daugintis, Rapolas and Barumerli, Roberto and Geronazzo, Michele and Pauwels, Johan and Picinali, Lorenzo and Poole, Katarina C.},
  journal = {IEEE Open Journal of Signal Processing},
  volume = {6},
  pages = {950--964},
  year = {2025},
}

\end{document}